# Thermalization of a SQUID chip at cryogenic temperature: Thermal conductance measurement for GE 7031 Varnish Glue, Apiezon N Grease and Rubber Cement between 20 and 200 mK

M. D'Andrea [*,1] • G. Torrioli[2] • C. Macculi[1] • M. Kiviranta[3]

[*]*Corresponding Author, matteo.dandrea@inaf.it*
[1]*INAF/IAPS, Via del Fosso del Cavaliere 100, 00133 Roma, Italy*
[2]*CNR/IFN Roma, Via del Fosso del Cavaliere 100, 00133 Roma, Italy*
[3]*VTT, Tietotie 3, 02150 Espoo, Finland*

**Abstract** In the context of the ATHENA X-IFU Cryogenic AntiCoincidence Detector (CryoAC) development, we have studied the thermalization properties of a 2mm x 2mm SQUID chip. The chip is glued on a front-end PCB and operated on the cold stage of a dilution refrigerator ($T_{BASE}$ < 20 mK). We performed thermal conductance measurements by using different materials to glue the SQUID chip on the PCB. These have been repeated in subsequent cryostat runs, to highlight degradation effects due to thermal cycles. Here, we present the results obtained by glues and greases widely used in cryogenic environments, i.e. GE 7031 Varnish Glue, Apiezon N Grease and Rubber Cement.

M. D'Andrea • G. Torrioli • C. Macculi • M. Kiviranta

## 1 Introduction

We are developing the Cryogenic AntiCoincidence detector (CryoAC [1][2]) of the ATHENA X-ray observatory [3]. It is a particle detector aimed to reduce the background of the X-IFU, the on-board X-ray spectrometer [4]. The CryoAC is a 4-pixels silicon detector, sensed by Transition Edge Sensors (TES) and readout by four independent single-stage SQUIDs operated in Flux-Locked-Loop (FLL) mode. The SQUID chips are glued on the detector CFEE PCB (Cold Front End Electronics Printed Circuit Board) located in the cold stage, at T = 50 mK. The CryoAC assembly concept is shown in Fig. 1.

The selection of the glue to be used on the PCB is an ongoing task managed at X-IFU system level, and it is driven by several constraints such as thermal properties, mechanical behavior and chemical composition. The proper glues qualification will be carried out inside the next CryoAC industrial contract (around 2025), and it will follow the standards for space product assurance (see [5] for detail about adhesive bonding qualification for space applications).
In this context, we have preliminarily investigated the thermal properties of different glues and greases widely use in cryogenic environments, in order to have a first reference for the CryoAC development activity.

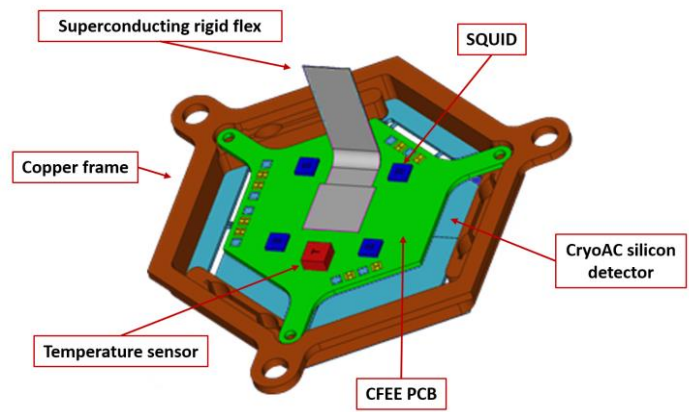

**Fig. 1** The Cryogenic Anticoincidence detector (CryoAC) cold stage (50 mK) assembly concept. (Color figure online)

## 2 Experimental Setup

We have directly studied the thermalization of a SQUID chip glued on a CFEE PCB prototype featuring superconducting Al traces (designed by us and

# Thermalization of a chip SQUID at cold

provided by Omni Circuit Boards[1]). The system has been anchored on the cold stage of a dilution refrigerator, with a base temperature $T_{BASE} < 20$ mK.

The SQUID chip features two very low-ohmic embedded resistors, usually used to shunt the CryoAC TES. We connected one of these to the SQUID input coil, to use the device as a Johnson noise thermometer. By varying the SQUID working point, we measured the actual chip temperature for different power loads (due to the SQUID bias) and different cold stage temperatures, effectively measuring the thermal conductance between the chip and the thermal bath. The latter is represented by the PCB copper ground plane, which has been strongly coupled to the mixing chamber via screws and gold bonding wires.

We have performed the measurements by using different materials to glue the SQUID on the PCB, i.e. GE 7031 Varnish Glue[2], Apiezon N Grease[3] and Rubber Cement[4]. Furthermore, for each material we have repeated the measurements in two subsequent cryostat run, to highlight degradation effect due to thermal cycle.

A picture and a thermal schematic of the setup are shown in Fig. 2.

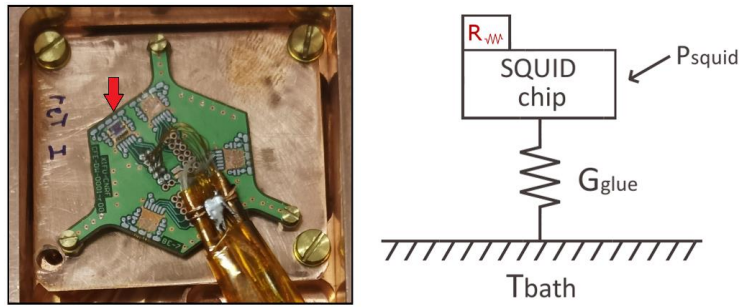

**Fig. 2** *Left* Picture of the experimental setup, showing the SQUID chip (red arrow) glued on the CFEE PCB prototype. *Right* Thermal model of the system. The resistor "R" is embedded on the SQUID chip. $P_{SQUID}$ is given by the SQUID bias and $T_{bath}$ is the temperature of PCB copper ground plane, strongly anchored to the cryostat cold stage. $G_{glue}$ is the thermal conductance of the glue used to bond the SQUID chip to the PCB. (Color figure online)

---

[1] https://www.omnicircuitboards.com/
[2] Commercialized as "IMI 7031 insulating varnish" from GVL Cryoengineering, http://www.gvl-cryoengineering.de/
[3] https://apiezon.com/products/vacuum-greases/apiezon-n-grease/
[4] Commercialized as "PAX Rubber Cement (Solution)", https://pax.com.tw/

M. D'Andrea • G. Torrioli • C. Macculi • M. Kiviranta

*2.1 SQUID schematics and characteristics*

The SQUID used for the measurements is a series array model M4A produced by VTT [6]. It has been deposited on a 2 mm x 2 mm silicon chip (675 μm thick), were only the top surface has been polished. Its schematics, both in term of circuitry and pads layout, are shown in Fig. 3.

In addition to the standard circuitry for SQUID operations (i.e. input and feedback coils), the chip features two very low-ohmic resistors, with $R_1$ = 0.5 mΩ and $R_2$ = 2.2 mΩ nominal values. Here, the 0.5 mΩ resistor has been directly connected to the input coil via aluminum bonding wires, to use the device as Johnson noise thermometer (red connections in Fig. 3 *Left*). The chip features also an embedded heater, which has not been used for this application.

The SQUID has been operated by a commercial Magnicon XXF-1 electronics[5]. The characteristics V-Φ curves and the measured value of its main parameters are shown in Fig. 4 . Around the maximum gain working point ($I_B$ ~ 12.0 μA, $V_{SQUID}$ ~ 200 μV, dV/dΦ ~ 3 mV/$\Phi_0$), the SQUID has a typical power dissipation $P_{SQUID}$ ~ 2.4 nW.

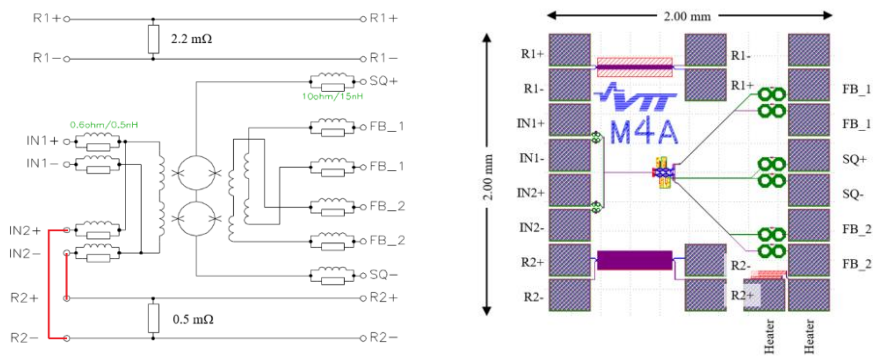

**Fig. 3** VTT SQUID series array M4A schematics. *Left* Simplified internal schematics. For these measurements, the R2 resistor has been directly connected to the SQUID input coil (red lines). *Right* Pad layout and chip size. (Color figure online)

---

[5] http://www.magnicon.com/squid-electronics/xxf-1

## Thermalization of a chip SQUID at cold

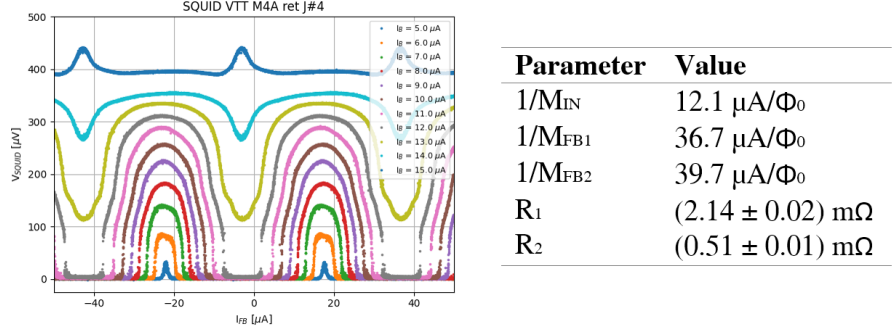

| Parameter | Value |
| --- | --- |
| $1/M_{IN}$ | 12.1 µA/$\Phi_0$ |
| $1/M_{FB1}$ | 36.7 µA/$\Phi_0$ |
| $1/M_{FB2}$ | 39.7 µA/$\Phi_0$ |
| $R_1$ | (2.14 ± 0.02) mΩ |
| $R_2$ | (0.51 ± 0.01) mΩ |

**Fig. 4** *Left* Characteristics V-Φ curves of the SQUID acquired at 50 mK through the feedback line, for different bias currents. *Right* Measured values of the main SQUID parameters. $1/M_{IN}$, $1/M_{FB1}$ and $1/M_{FB2}$ are the input and feedbacks mutual inductance, $R_1$ and $R_2$ the embedded resistors. (Color figure online)

## 3 Thermal conductance measurements

To measure the thermal conductance between SQUID and PCB ($G_{GLUE}$), we monitored the chip temperature ($T_{SQUID}$) for different power loads ($P_{SQUID}$) and thermal bath temperatures ($T_B$).

The power loads on the SQUID is set by its bias point, and it has been evaluated from the acquired characteristic curves (for each bias point, $P_{SQUID} = I_B \cdot V_{SQUID}$ ). The chip temperature has been evaluated by measuring the white noise at the SQUID output ($W_n$ [$\Phi_0$/√Hz]), which is due to the Johnson noise of the resistor connected to the input coil ($I_n$ [A/√Hz]) [7]:

$$W_n = \frac{1}{M_{IN}} \cdot I_n = \frac{1}{M_{IN}} \sqrt{(4 \cdot k_B \cdot T_{SQUID})/R_2} \quad (1)$$

$$T_{SQUID} = \left(W_n \cdot \frac{1}{M_{IN}}\right)^2 \cdot \frac{R_2}{4 \cdot k_B} \quad (2)$$

where $1/M_{IN}$ is the input coil coupling, $R_2 = 0.5$ mΩ and $k_B$ is the Boltzmann constant. Fig. 5 *Left* shows the noise spectra acquired at a fixed bath temperature for different SQUID power loads (i.e. different SQUID bias point). $W_n$ has been evaluated as the average noise value in the 200 Hz - 3 kHz frequency range. Note that the typical total system noise at open input (blue line in Fig. 5 *Left),* mainly due to the electronics pre-amplifier and previously measured, is negligible with respect to the Johnson noise.

Therefore, we are confident that the measured noise is given by the $R_2$ Johnson component.

To evaluate the glue thermal conductance, the acquired data have been finally fitted with the standard power-law model describing the power flow to the heat bath [8]:

$$P_{SQUID} = k \cdot (T_{SQUID}^n - T_B^n) \qquad (3)$$

where k and n are the fit parameters, related the nature of the thermal link. The final measured thermal conductance is:

$$G_{GLUE}(T) = n \cdot k \cdot T^{(n-1)} \qquad (4)$$

One of the performed measurements is presented in Fig. 5 *Right* (see the Appendix A for the complete set of measurements). Note that, due to the non-ideal thermal coupling between the SQUID chip and the PCB (which is strongly anchored to the bath), at low temperature the chip is significantly warmer than the bath temperature.

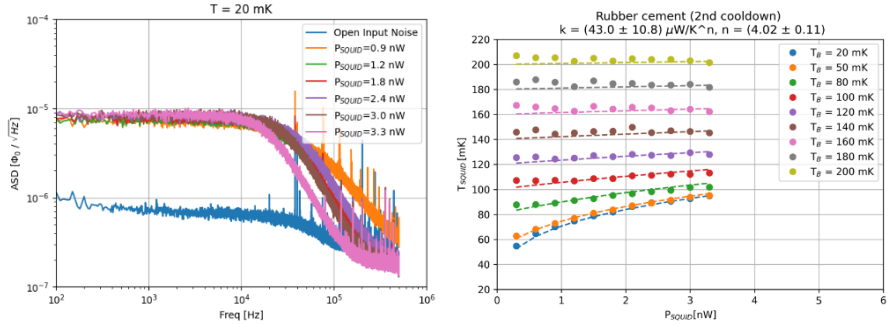

**Fig. 5** *Left* Noise spectra acquired at a fixed bath temperature ($T_B$ = 20 mK), used to evaluate the chip temperature as a function of the power load. The blue curve represent the total system noise at open input, previously measured. (Color figure online). *Right* Thermal conductance measurement. The SQUID chip temperature (evaluated from Johnson noise) is reported as function of power load, for different bath temperatures. Data has been fitted with eq. (3) (dashed lines). (Color figure online)

# Thermalization of a chip SQUID at cold

*3.1 Control measurement*

We performed also a control measurement to check potential systematic setup effects. In this case, we measured the temperature of a second SQUID chip (with embedded resistor) directly glued to the copper plate hosting the PCB, i.e. to the thermal bath. Also here, we performed Johnson noise thermometry by connecting only the resistor of the second chip to the PCB SQUID input coil (via aluminum bonding wires). A picture and a thermal schematic of the control setup are shown in Fig. 6.

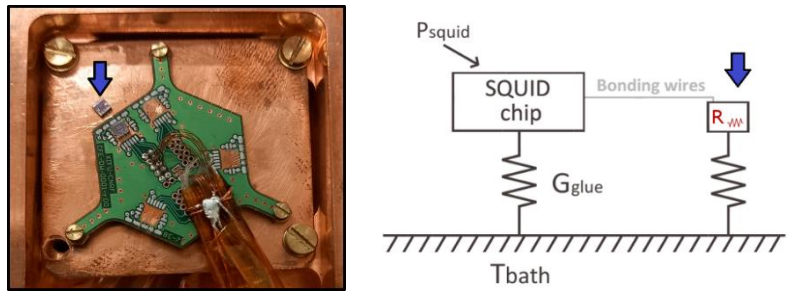

**Fig. 6** *Left* Picture of the control measurement setup. A second chip (blue arrow) is glued directly to the thermal bath. Its embedded resistor is connected to the PCB SQUID via Al bonding wires, to perform Johnson noise thermometry. *Right* Thermal model of the control setup. Note that the second chip (blue arrow) has not direct power loads. (Color figure online)

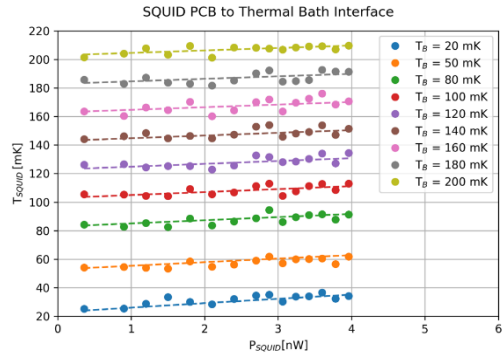

**Fig. 7** Results of the control measurements. The temperature of the second chip is monitored as a function of the PCB SQUID power, for different thermal bath temperature. (Color figure online)

The result of the control measurement is reported in Fig. 7. In this case, we found a good thermalization of the chip, with its temperature always close to the thermal bath one, especially at low power values. This is expected, since in this configuration the monitored chip has not direct power loads. The slight dependence of the chip temperature from the PCB SQUID power is probably due to parasitic heat load, mainly due to the aluminum bonding wires.

This measurement shows that no significant systematic effect affects the setup, and that it is possible perform a proper Johnson noise thermometry down to ~ 20- 30 mK by this setup. This is important to exclude, for example, possible effects due to the electron-phonon decoupling in the monitored resistor. The electron-phonon conductance has indeed a strong dependence from the temperature, with thermal power index n ~ 5-6 [8][9], thus it could be a limiting factor for measurements at ultra-low temperature.
Furthermore, the measurements guarantees that all the additional component in the noise (i.e. SQUID, pre-amplifier and DAQ noise) are effectively negligible with respect to the Johnson noise of the resistor.

## 4 Results

The results of the measurements performed with the different glues are summarized in Fig. 8 and Tab. 1. For each material, they are reported the best-fit k and n parameters and the resulting thermal conductances (evaluated at different temperatures) in two subsequent cooldowns.
The gluing of the SQUID chip has been always performed by hand, in air, without any curing procedure. The bond line thickness was of the order of 50 µm, roughly controlled by visual inspection. Air bubbles were removed from the bonding agents before placing the chip on the PCB. In each cryostat run, the setup undergo a vacuum pump-down (from 1000 mbar to < $10^{-4}$ mbar) and a full thermal cycle from 300 K to 20 mK.

We observe that GE 7031 shows the worst thermal conductance at cold, and a significant degradation with thermal cycle. Rubber Cement and Apiezon N grease show similar thermal properties, without significant degradation in the $2^{nd}$ cooldown and with G of the order of 150 nW/K @100 mK.
Note that the best-fit power index n is always compatible with the value n = 4, as expected for a Kapitza (i.e. boundary) interface [9].

## 5 Conclusions

In the context of the ATHENA X-IFU Cryogenic AntiCoincidence detector (CryoAC) development, we have studied the thermalization properties of a 2 mm x 2 mm SQUID chip bonded on a PCB with different gluing materials. In Tab. 1 and Fig. 8 are reported the measured thermal conductances for GE-7031 Varnish Glue, Apiezon N Grease, and Rubber Cement between 20 mK and 200 mK, in two subsequent thermal cycles. These values represent a first

## Thermalization of a chip SQUID at cold

reference for the CryoAC development activity and worth of notice for the whole Low Temperature Detector community.

Finally, we report that the next-generation CryoAC SQUIDs, currently under development at VTT, will include also a gold bonding pad. This will allow, if needed, to further improve the SQUID chip thermalization via gold bonding wires.

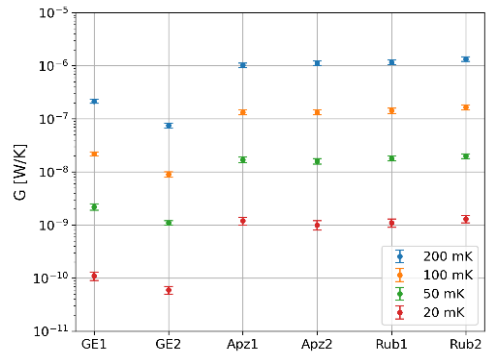

**Fig. 8** Summary of the Thermal Conductance measurements performed with different materials (GE-7031 Varnish Glue, Apiezon N Grease and Rubber Cement) in subsequent thermal cycles. (Color figure online)

| Dataset | k [μW/K] | n | G @ 50 mK [nW/K] | G @ 100mK [nW/K] |
|---|---|---|---|---|
| GE-7031 (1st cooldown) | 10.2 ± 1.1 | 4.30 ± 0.06 | 2.2 ± 0.3 | 22 ± 2 |
| GE-7031 (2nd cooldown) | 2.7 ± 0.2 | 4.09 ± 0.04 | 1.1 ± 0.1 | 9 ± 1 |
| Apiezon N (1st cooldown) | 29.9 ± 6.8 | 3.95 ± 0.10 | 17 ± 2 | 133 ± 15 |
| Apiezon N (2nd cooldown) | 38.5 ± 7.7 | 4.07 ± 0.09 | 16 ± 2 | 133 ± 15 |
| Rubber Cement (1st cooldown) | 37.3 ± 7.8 | 4.02 ± 0.09 | 18 ± 2 | 143 ± 16 |
| Rubber Cement (2nd cooldown) | 43.0 ± 10.8 | 4.02 ± 0.11 | 20 ± 2 | 165 ± 18 |

**Tab. 1** Summary of the Thermal Conductance measurements performed with different materials and in subsequent thermal cycles.

M. D'Andrea • G. Torrioli • C. Macculi • M. Kiviranta

**Appendix A**

The full set of thermal conductance measurements performed is reported in Fig. 9.

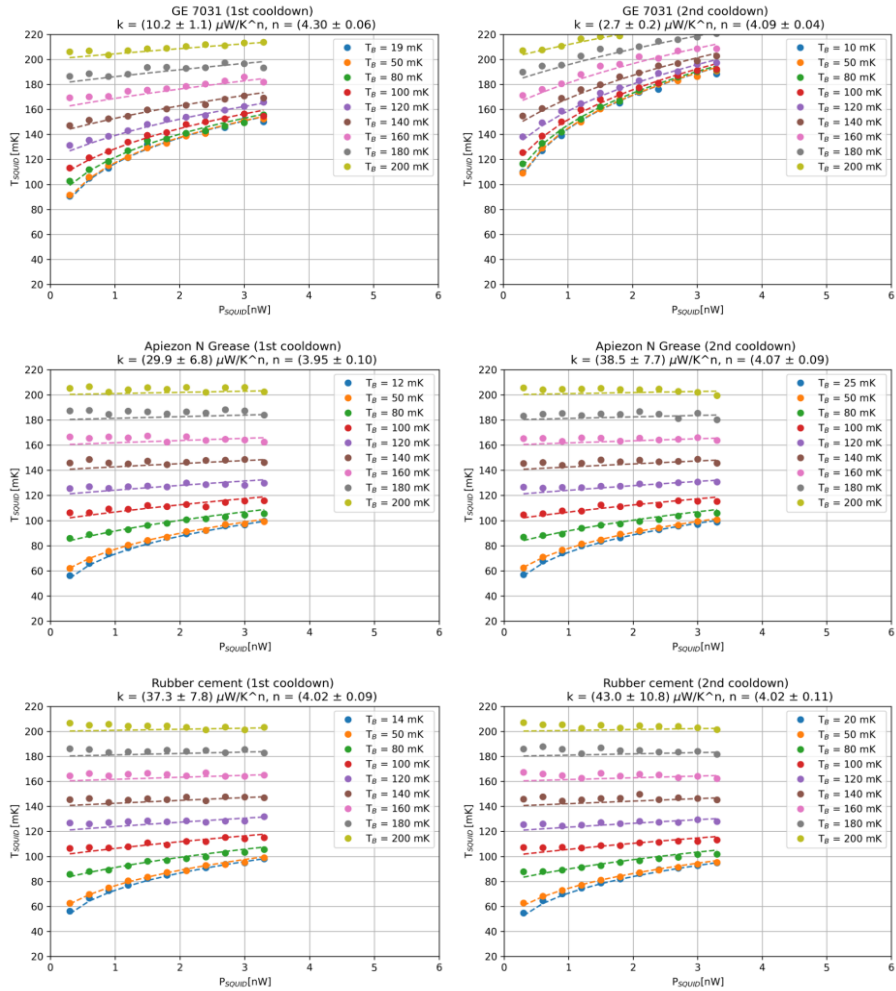

**Fig. 9** Thermal Conductance measurements ($T_{SQUID}$ vs $P_{SQUID}$ for different thermal bath temperatures) performed for different gluing materials (*Top* GE 7031, *Middle* Apiezon N Grease, *Bottom* Rubber Cement) in two subsequent cryostat runs (*Left* 1st cooldown, *Right* 2nd cooldown). Setup as from Fig. 2. (Color figure online)

# Thermalization of a chip SQUID at cold

**Acknowledgements** This work has been supported by ASI (Italian Space Agency) contract n. 2019-27-HH.0. The authors would like to thank Jan van der Kuur (SRON) for useful discussion. M.D. thanks Ciclofisica (the Ciclofficina «Roberto Perciballi») for the Rubber Cement supply.